# Transcriptome profiling research in urothelial cell carcinoma

Umar Ahmad[1,2*], Buhari Ibrahim[3,4], Mustapha Mohammed[5,6], Ahmed Faris Aldoghachi[7], Mahmood Usman[8,9], Abdulbasit Haliru Yakubu[10,11], Abubakar Sadiq Tanko[12], Khadijat Abubakar Bobbo[13,14], Usman Adamu Garkuwa[15,4], Abdullahi Adamu Faggo[16], Sagir Mustapha[17], Mahmoud Al-Masaeed[18,19], Syahril Abdullah[13,1*], Yong Yoke Keong[20], Abhi Veerakumarasivam[21]

[1]Medical Genetics Laboratory, Genetics and Regenerative Medicine Research Centre, Faculty of Medicine and Health Sciences, Universiti Putra Malaysia, Serdang, Selangor, 43400, Malaysia, [2]Medical Genetics Unit,, Department of Anatomy, Faculty of Basic Medical Science, Bauchi State University, Gadau, Bauchi, PMB 65, Nigeria, [3]Department of Imaging, Faculty of Medicine and Health Sciences, Universiti Putra Malaysia, Serdang, Selangor, 43400, Malaysia, [4]Department of Physiology, Faculty of Basic Medical Sciences, Bauchi State University, Gadau, Bauchi, PMB 65, Nigeria, [5]School of Pharmaceutical Sciences, Universiti Sains Malaysia, Penang, Pulau Pinang, 11800, Malaysia, [6]Department of Clinical Pharmacy and Pharmacy Practice, Faculty of Pharmaceutical Sciences, Ahmadu Bello University, Zaria, Kaduna, PMB 1045, Nigeria, [7]Medical Genetics Laboratory, Department of Biomedical Science, Faculty of Medicine and Health Sciences, Universiti Putra Malaysia, Serdang, Selangor, 43400, Malaysia, [8]Department of Human Anatomy, Faculty of Basic Medical Sciences, Ahmadu Bello University, Zaria, Kaduna, PMB 1045, Nigeria, [9]Department of Human Anatomy, Faculty of Basic Medical Sciences, Yusuf Maitama Sule University, Kano, PMB 3220, Nigeria, [10]Faculty of Pharmacy, University of Maiduguri, Maiduguri, Borno, PMB 1069, Nigeria, [11]Department of Pharmaceutical Service, University of Maiduguri Teaching Hospital, Maiduguri, Borno, PMB 1414, Nigeria, [12]Department of Biochemistry, Faculty of Science, Bauchi State University, Gadau, Bauchi, PMB 65, Nigeria, [13]UPM-MAKANA Cancer Research Laboratory, Institute of Bioscience, Universiti Putra Malaysia, Serdang, Selangor, 43400, Malaysia, [14]Department of Anatomy, Faculty of Basic Medical Sciences, Gombe State University, Gombe, PMB 127, Nigeria, [15]Department of Human Physiology, Faculty of Basic Medical Sciences, Ahmadu Bello University, Zaria, Kaduna, PMB 1045, Nigeria, [16]Department of Microbiology, Faculty of Science, Bauchi State University, Gadau, Bauchi, PMB 65, Nigeria, [17]Department of Pharmacology, School of Medical Sciences, Universiti Sains Malaysia, Kota Bharu, Kelantan, 16150, Malaysia, [18]Department of Nursing, Faculty of Health and Medicine, University of Newcastle, Callaghan, NSW, 2308, Australia, [19]Faculty of Medicine and Health Sciences, Universiti Putra Malaysia, Serdang, Selangor, 43400, Malaysia, [20]Department of Human Anatomy, Faculty of Medicine and Health Sciences, Universiti Putra Malaysia, Serdang, Selangor, 43400, Malaysia, [21]Department of Medical Sciences, School of Science and Technology, Sunway University, Bandar Sunway, Selangor, 47500, Malaysia

[*]Corresponding authors: umarahmad@basug.edu.ng, syahril@upm.edu.my

**Abstract**

Urothelial cell carcinoma (UCC) is the ninth most common cancer that accounts for 4.7% of all the new cancer cases globally. UCC development and progression are due to complex and stochastic genetic programmes. To study the cascades of molecular events underlying the poor prognosis that may lead to limited treatment options for advanced disease and resistance to conventional therapies in UCC, transcriptomics technology (RNA-Seq), a method of analysing the RNA content of a sample using modern high-throughput sequencing platforms has been employed. Here we review the principles of RNA-Seq technology and summarize recent studies on human bladder cancer that employed this technique to unravel the pathogenesis of the disease, identify biomarkers, discover pathways and classify the disease state. We list the commonly used computational platforms and software that are publicly available for RNA-Seq analysis. Moreover, we discussed the future perspectives for RNA-Seq studies on bladder cancer and recommend the application of new technology called single cell sequencing (scRNA-Seq) to further understand the disease.

## 1.0 Introduction

Cancer represents one of the most life-threatening diseases posing a major health challenge all over the world (1, 2). Urothelial cell carcinoma (UCC) also known as bladder cancer is among the principal causes of cancer-related deaths globally (3, 4). The incidence of bladder cancer is ranked $9^{th}$ with a total number of 549,393 new cases and 199,922 deaths globally (5, 6). These numbers are forecasted to double in 2040, particularly in the developed countries, signifying a severe health crisis in the future (5, 7) [,7] which can lead to financial burden on countries due to the exorbitant treatments, high recurrence rates and subsequent need for long-term follow-up (8). Urothelial cell carcinoma is reported to be the second most frequent malignant tumour of the genitourinary tract with men at risk four times higher than women.[4] The most common bladder cancer is transitional cell carcinoma (TCC) or urothelial cell carcinoma (UCC), classified into; non-muscle-invasive bladder cancers (NMIBCs) and muscle-invasive bladder cancers (MIBCs). The NMIBCs accounts for about 70-80 per cent of all diagnosed patients, with the tendency of future recurrences and may advance into muscle-invasive bladder cancers (9). Whereas MIBCs accounts for about 20-25 per cent of patients, signifying an active, locally invasive carcinoma with a metastatic potential (10).

Despite diagnostic and therapeutic advances for UCC, if it is not detected early, will remain a big challenge. Although there have been great achievements in its management and treatments that include surgical procedures, radiotherapy, pre- and post-operative treatments and chemotherapy , there has been no significant progress in survival rates for bladder cancer patients (11). Moreover, these treatments are presented with various side effects that lead to other health problems (4). As a result, healthcare providers receive substantial number of cases with relapse and progression, leading to long-term follow-up. All these challenges made bladder cancer an expensive disease to manage (7). The need to identify novel molecular markers in bladder cancer in order to predict medical outcomes, especially in patients with relapse has made researchers in recent years to focus more on the molecular aspects of the disease. This development came with the aim of boosting its biological understanding in order to unravel the molecular factors that can be utilized as targets for therapies and guiding assessment of risk and help in informed decision making in clinical practice (12).

Transcriptomic technologies employ methods that study and quantify organism's transcriptome (13), a complete set of RNA transcripts in a cell for a specific developmental stage or physiological state. Understanding transcriptome is essential for interpreting the functional elements of the genome and revealing the molecular constituents of the cells and tissues to better understand the development of diseases (14). Identifying transcripts and quantifying the level of gene expression have been the major focus in molecular biology since the discovery of the RNA as an essential intermediate between the genome and the proteome (15). The last decades have witnessed the use of microarray technology and bioinformatic analysis in screening genetic changes at the genome level which has greatly assisted in the identification of differentially expressed genes (DEGs) and functional pathways implicated in cancer (16). However, with technological advancement and elucidation of noncoding RNAs, transcriptomic approaches have given room for deeper understanding of the intricacies of the regulation of gene expression, alternate splicing events, functions of noncoding RNAs and ascertaining the importance of such approaches in the correct construction and annotation of complex genomes (1).

Furthermore, high-throughput RNA sequencing (RNA-Seq) has opened opportunities for transcriptomic studies and has advanced into a standard technique used in biomedical studies. This technique is currently utilized in the estimation of gene expression, identification of noncoding genes, discovery of new genomic characteristics and drug discoveries. It is well established that RNA-Seq has strong advantages over the previously developed sequencing techniques (17). This makes RNA-Seq a vital tool in cancer biology that can be exploited for tumour classification, patient stratification and monitoring of patients' response to therapy (18). However, despite the high level of correlation that exists between RNA-Seq and other techniques such as microarray, studies have strongly emphasised the merits of RNA-Seq over the other techniques (19). The discovery of gene fusion and differential expression of RNA transcripts that are known for causing diseases are some of the prospects of the RNA-Seq (20). In this review, we summarized the current literature status of transcriptomic studies carried out on urothelial cell carcinoma and identified the possible areas for future research.

## 2.0 Transcriptome sequencing technologies

Transcriptome sequencing technology is a computational method that determines the identity and abundance of RNA sequences in a biological sample. This technology has been widely used to study dynamic gene expression in diverse human tissues, including human bladder cancer. The general workflow of RNA-Seq begins with isolation of RNAs from tissue samples, library preparation, sequencing and computational analysis (**Figure 1**). Below we summarize and discuss RNA-Seq techniques, narrowing down to its application in human bladder cancer studies.

### 2.1 Principle of RNA-Seq technology

RNA-Seq is a transcriptome profiling technology that utilizes next-generation sequencing (NGS) platforms (21). The principle of RNA-Seq involve the reverse-transcription of RNA-Seq transcripts into cDNA, ligation of adapters to each end of the cDNA, sequencing and then aligning them to a reference genome or assembling to obtain *de novo* transcripts, proving a genome-wide expression profile (22). The quality and quantity of the starting RNA material are the most important aspects to consider when deciding on the methods to generate RNA-Seq libraries (23, 24). While working with RNA, it is critical to avoid RNase contamination by using sterile, RNase-free solutions, and plastic ware. It is also important to include quality control (QC) in all stages of operation (23). The method of RNA isolation affects the ability to detect differentially expressed transcripts (25). Various methods of RNA extraction exist in the literature; the most popular and widely used methods are the TRIzol (Life Technologies) and RNAeasy (Qiagen) (25). Other methods include density gradient centrifugation, magnetic bead technology, lithium chloride and urea isolation, Oligo(dt)-cellulose column chromatography, and non-column poly (A)+ purification/isolation (26). TRIzol employs the guanidinium-acid-phenol extraction and regarded as the 'gold standard' but chemicals used are corrosive and toxic, whilst, the RNAeasy (and variants) employs the glass fibre filter and silica technology that is fast, simple and safe to use but expensive (24, 25).

The extracted RNA is treated in solution or on-column with DNase to remove traces of genomic DNA and to prevent contamination of RNA-Seq libraries by DNA. DNA-free RNA could then be assessed for both quality and quantity (24). In assessing the quality of the RNA, most laboratories utilize the electrophoretic-based system (22, 24). The quality of the RNA produced is measured through the RNA integrity score, RNA Integrity Numbers (RIN) (23, 24). The measure reads per kilobase of exon model per million reads (RPKM) and its variance are the most frequently adopted in the measurement of RNA-Seq expression value (27). The acquisition of RNA-Seq data consists of several steps and in each of these steps, specific QC checks are applied to monitor the quality of the data output produced (15).

## 2.2 RNA-Seq library preparation

In the last decade, there has been a considerable understanding of the dynamic nature of organism's genome (13, 28), this is not limited to the modifications of DNA sequence but also to the quantitative and qualitative measurements of RNA sequences (13). Before sequencing a sample, a library must be prepared for that sample (13, 28, 29). A library is a collection of randomly sized DNA fragments representing the sample input (28). Depending on the type of NGS application, different library preparation steps are available (28-30). Also, the sample preparation required for RNA sample sequencing vary depending on the type of RNA (28, 30, 31). Steps that are required for RNA-Seq are; selection of transcript with poly(A) tail, rRNA depletion (28, 30), fragmentation followed by cDNA synthesis, purification and amplification (30, 32). mRNA and lncRNA >200 nt contains a poly(A) tail (28, 29), convenient for the enrichment of poly(A) + RNAs from total cellular RNA, which is carried out with oligo-dT nuclease (28, 29). This is followed by rRNA depletion (32), this can be done based on sequence-specific probes which can hybridize to rRNA then depleted with streptavidin beads (28). RNA samples are fragmented to a certain size range before reverse transcriptase (RT) (32), this is necessary due to size limitation of most sequencing platforms (28). Fragmentation of RNA can be done with alkaline solution or divalent cations on an elevated temperature (28). Enzymes such as RNase III can also be used to fragment RNAs but this can introduce bias due to its preference for double-stranded RNA sequence (28). cDNA synthesized from the RNA fragments using RT are ligated to DNA adapters before amplification and sequencing (28, 29, 32).

### 2.3 Sequencing platforms

The method that has been in place for the past few decades is Sanger sequencing, this has been characterized by its simplicity but is capital intensive and the process takes longer time than necessary (33). Novel sequencing technologies that immediately evolved after Sanger sequencing are collectively termed as next-generation sequencing (NGS) and are often described as throughput, accurate, cost-effective and reliable techniques that can examine the whole genome within short period of time (33, 34). Generally, NGS platforms are grouped into second and third generations. The most widely used platforms for second generation sequencing are Ion Torrent and Illumina, while Pacific Biosystems and Oxford Nanopore Technologies are the most popular platforms for third generation sequencing (35). Also, 454 FLX is among the second generation platforms that was released in 2005 and works based on pyrosequencing (i.e. sequencing-by-synthesis technique) like Illumina (36). The most established Illumina platforms include MiSeq, HiSeqs(37) and NovaSeq, the newest platform that is used for large-scale whole genome sequencing analysis (38). MiSeq works as personalized sequencer that sequence very small genomes with high speed and complete its operation within 4 hours. On the other hand, HiSeqs such as HiSeq 2500 is designed for "high-throughput" usage and mostly completes its cycle in 6 days period (37). Unlike Illumina and 454 sequencing platforms that are sequenced by synthesis, sequencing by oligonucleotide ligation and detection (SOLiD) developed by Applied Biosystems (which later became Life Technologies) is by ligation through hybridization of short probes with the template DNA strand (36). Despite its apparent advantages, its reads length and depth are not as sophisticated as that of Illumina, thereby causing major assembly challenge (39, 40). Throughout all Illumina platforms, only 1% error rate is recorded and substitution is regarded as their major error (37), providing in depth sequencing capacity that allows detection of very small quantity of transcripts in the sample (41).

Nevertheless, the third NGS sequencing platforms provide two essential advantages over the second generation, including their ability to increase reads length, avoid partiality of PCR in the amplification process and the capacity to sequence single molecules at a given time (36). Despite all advantages of NGS, the technique has faced numerous challenges including numerous GC content, big size of genome and the presence of homopolymers. However, various alternatives have been put in place to overcome the current challenges (28). Moreover, third NGS sequencing platforms such as single-molecule real-time (SMRT) sequencing that was developed by Pacific Biosciences (PacBio), uses long reads that provides solution to the myriads challenges faced by second generation platforms due to their short reads length that make them unable to accurately detect gene isoform, poor genome assembly and resolution of complicated genomic region. On the contrary, PacBio sequencing (SMRT) is limited by high percentage of error, high cost per base and ~~has~~ low-throughput (42). These challenges can be overcome by the use of platforms with lower error rates as low as 3%(38) , and alternatively the use of hybrid Sanger sequencing and PacBio sequencing technology have been proposed (42). In addition, Helicos is one of the single-molecule-based platforms with 5% error rate, reducing the number of usable reads, making the reference genome extremely hard to be matched with the sequence reads, and causing the loss of miRNA reads at the alignment level (41). Due to their low error rate (>1%), Illumina or SOLiD platforms are regarded as the best alternative for miRNA sequencing because of its small size (41). Nanopore sequencing platforms such as MinION, PromethION and GridION(37) belong to third generation sequencing platforms developed by Oxford Nanopore Technologies, which operates by passing a single stranded molecule of DNA or RNA via a protein nanopore at the rate of 30 bases per second, through an electrical current allowing direct sequencing of the molecule, thus offering greater advantages (43). However, a major setback of these platforms is high error rate up to 10% to 20%, when compare to other platforms with high-throughput (37, 44).

Most recently, several studies have reported remarkable achievements in the diagnosis of different cancers including bladder cancer using RNA-Seq (45-47). Transcriptomics (RNA-Seq) refers to the application of any NGS platforms for the examination of RNA(41) and has become the best method for whole-transcriptome profiling over the last decade (48). The selection criteria for each platform depend on the purpose of the experiments. The techniques operate similar to DNA sequencing except for the library preparation and their data analysis which comprises assembly of transcript, uncovering novel transcripts and calculation of transcripts, among others (41). Additionally, RNA-sequencing gives comprehensive, high-throughput, precise, accurate and impartial view of the transcriptome analysis that overcome the inherent limitations of real time PCR and microarray techniques (49). These limitations include: the need for previous knowledge of the sequence in question, inability to calculate low and high expressed genes with great accuracy, among others (22). Although RNA-Seq is high- throughput, it is too expensive (50), especially for clinical settings.

## 3.0 RNA-Seq analysis pipelines

Over the years, microarray and gene-chip technologies provide an insight into understanding the genetic changes in biological samples. However, these techniques are known to have certain limitations related to dynamic range, resolution and accuracy (51) Advances in transcriptome technology have allowed deeper understanding of the intricacies of gene expression regulation, particularly high-throughput RNA sequencing technology that made it possible to observe whole transcriptome variations, discover novel splicing sites and events, functions of noncoding RNAs, as well as proving correct construction and annotation of complex genomes (52). It also aids to qualitatively ascertain the RNA transcripts present, RNA editing sites, and to quantitatively know how much of the individual transcripts are being expressed (53). Thus, it is paramount to overview pipelines and workflows applied to bladder cancer RNA-Seq analyses.

A number of computational pipelines and workflows are being used for the pre-processing of RNA-Seq data in cancer studies and other experimental purposes (50, 54, 55). A typical RNA-Seq workflow consists of seven steps; (1) pre-processing of raw data, (2) alignment of reads to the reference, (3) transcriptome reconstruction, (4) quantifications of transcripts or genes level, (5) differential expression analysis, (6) functional profiling, and (7) advanced analysis (**Figure 2**) (56). These stages in the RNA-Seq workflow that includes quality control (QC) and data analysis can be done using varieties of computational platforms or tools. For example, read counts may be aligned using different tools such as spliced transcript alignment to a reference (STAR) or Tophat (57, 58). Then, the aligned read counts can be obtained using either HTSeq or Rsubread R/Bioconductor package (59, 60). The advantage of Rsubread over HTSeq is that the former is faster, requires less memory and summarizes the read counts that are more closely related to a true value (61, 62).

RNA-Seq raw data often have quality problems that can distort analytical findings significantly and lead to incorrect conclusions (63). For instance, the quality of raw RNA-Seq data could be altered by residue of ribosomal RNA, degradation of RNA and variation in read coverage (63). Hence, in order to obtain accurate transcripts or genes measurements and proper acquisition of information from the data, raw RNA-Seq data must be reviewed and evaluated by quality control measures before subsequent analyses are conducted (27, 63). Presently, the most widely and commonly used computational tools available for RNA-Seq QC include; FASTQC and MultiQC. FASTQC processes one sample at a time, while MultiQC can generate a single report that visualizes the output of several samples from multiple tools, thereby giving room for easy comparison (64, 65). Other important and commonly used computational software for QC are RseQC, RNA-seQC and RNA-QC-Chain (66-68). Although both RseQC and RNAseQC can offer QC statistics of aligned read counts, RseQC partially relies on the University of California Senta Crus (UCSC) Genome Browser (67). Moreover, they are slow and unable to provide sequence trimming and filtration of contaminants. However, RNA-QC-Chain can remove low quality reads and contamination, in addition to providing fast and reliable QC to produce data for downstream analysis (63). RNA-Seq data analyses steps totally depend on the data quality and specific aims of the study. These analyses steps were reviewed in detail elsewhere (27, 69).

The system of RNA-Seq analysis employs high-computational tool applications for the development of pipelines that orchestrate the entire workflow and optimize usage of available computational resources (67). The development of such analytic tools for RNA-Seq data has expanded owing to complex nature of transcriptome data, and thus, selecting the correct processing pipeline and normalization strategy has a significant impact on downstream analysis (70). This pipeline consists of multiple independent analytical software packages, tools and platforms which employ R and Python, Unix/Bash, Java script, Perl and C++. Being that these software are in programmable environment; they provide flexible manipulation of data and methods. However, they required the user to have expertise in programming languages especially the bash language or Unix Commands Line (71). With the growing application of RNA-Seq in biomedical research, an integrated user-friendly platforms are needed to overcome the barriers encountered when using code-bond platforms. The Graphical user interface (GUI) or web-based platforms provide convenient and ~~enabling~~ (permissive?) environment for non-expert with advantages for quick exploratory analysis, even though not on the scale of large datasets (71). **Table 1** provides a summary of the various computational tools and their associated platforms used in RNA-Seq analyses.

Variations in the RNA-Seq analysis results might be observed due to usage of different platforms and analytical framework. The number of computational tools and bioinformatics methods that are currently in use add more challenges to the analysis and interpretation of the RNA-Seq data. In order to solve these challenges caused by variations in RNA-Seq analysis techniques, standard pipelines need to be enforced and re-designed in order to integrate analysis of multiple experiments. Workflow constructions software packages such as Chipster (72), Anduril (73, 74) and Galaxy (75) could be very much relevant in solving some of these challenges. For example, Anduril was developed for designing complex RNA-Seq pipelines with large-scale datasets, which require automated parallelization. While Chipster and Galaxy are powerful in data integrative visualization, making them very useful for data exploration and interpretation. Other workflows and management frameworks for RNA-Seq analysis are KNIME(76) which aid in visual assembly and interactive execution of data pipeline and Snakemake(77), which is a Python-based workflow management engine that provides a powerful execution environment. Workflow management framework that specifically focuses on RNA-Seq data analysis is reviewed by (78). In addition, the large-scale nature of the data analyses associated with RN-Seq brought many challenges that are beyond the scope of this review. Han and colleagues(78) reviewed these challenges comprehensively and proposed solutions. Moreover, results from RNA-Seq study on tumours revealed the presence of molecular subsets of cellular signatures, microenvironment and facilitates choices to circumvent treatment failure (79). Thus, single-cell sequencing (scRNA-Seq) may prove to be the correct method to understand tumour progression, pathogenesis and discovery of biomarkers that could lead to a better treatment and management of bladder cancer.

## 4.0 Review of RNA-Seq studies on bladder cancer

RNA sequencing has emerged as a powerful next-generation sequencing (NGS) technology tool for unbiased identification of gene expression, discovery and quantification of novel transcripts, and identification of alternatively spliced genes (22, 41). Consequently, recent progress in RNA-Seq has broadened the understanding of the molecular pathogenesis of cancers, including bladder cancer. RNA-Seq has been successfully applied in bladder cancer research for earlier detection, establishing pathological origin, and defining the aberrant genes and dysregulated molecular pathways across patient groups. Thus far, eight studies have utilized RNA-Seq technology in understanding the bladder cancer pathology. These studies are summarised based on sample, bladder cancer grades and library preparation techniques employed (**Table 2**).

## 4.1 Pathogenesis of bladder cancer

The aetiology and pathogenesis of bladder cancer are still less understood even though it has been characterized to have a high degree of malignancy and relapse even after surgery (80-82). Several findings suggest that in humans, microbiome could be one factor that can promote the development of cancers, including bladder cancer (83, 84). For example, *Campylobacter* genus, an opportunistic bacterium of the urinary tract was found to have pathogenic potential, as it was able to invade epithelial cells, produce toxins that inhibit NK cells cytotoxicity hence, promoting evasion of an immune response (83). This genus has the ability to generate a pro-inflammatory environment that supports tumour progression (83). The development of bladder cancer has also been reported to be highly correlated with abnormal expression of noncoding RNAs and protein-coding genes (82). Wang et al(82) constructed three-layer network of miRNA-lncRNA data from several microRNAs and long noncoding RNAs databases to calculate the topology attributes of nodes and concluded that *E2F1* and *E2F2* are important target genes of miRNA-93 while *AKT3* is an important target gene of miRNA-195 and that their dysregulation may be closely related to cell proliferation and apoptosis in bladder cancer. Similar findings reported that the dysregulation of lncRNA and circRNA are important in bladder cancer pathogenesis and progression (85).

## 4.2 Biomarkers for bladder cancer

Cancer biomarkers are biological molecules available in tissues, body fluids, or blood that helps in cancer prognosis, diagnosis, prediction of response to treatment, and monitoring disease progression (86, 87). To date, there have been several studies conducted to determine potential bladder cancer biomarkers in order to enhance the diagnostic accuracy. (87, 88) However, due to the false-positive and false-negative results, the use of these biomarkers has sparked clinical controversy (89). Nevertheless, utilizing the potential biomarkers to enhance the therapeutic surveillance and outcome has been investigated. A study on 105 NMIBC patients showed the alteration of certain genes that includes *TP53, PIK3CA, FGFR, TERT* promoter, *STAG2, ARIDIA*, and *KDM6A*. Among these genes, the *TERT* promoter accounted for 73% of alteration in NMBIC (90). While in MIBC, RNA-Seq analysis on high-grade bladder cancer (urine sample) revealed high expression of 15 genes such as *PLEKHS1, CP, WNT5A, RARRES1, MYBPC1, AR, ROBO1, SLC14A1, AKR1C2, FBLN1, IGFBP5, STEAP2, ENTPD5, GPD1L*, and *SYBU* (91). Recently, Sucularli (92) determined genes that are differentially expressed among bladder cancer (n=404) compared to the normal bladder (n=28) and found 559 genes were downregulated and 171 were upregulated . Six genes were associated with the patient's survival that includes *CDC20, PTTG1, PLK1, SFN, CCNB1,* and *BUB1B* (92).

Furthermore, Shen et al (93) conducted a study on cancer stem cells to determine the potential of *Sox4* as a biomarker for bladder cancer. Findings from the study showed a reduction in sphere formation and an elevation in the levels of aldehyde dehydrogenase in cells and the tumour forming ability upon the knockdown of *Sox4*. Moreover, the elevated expression of *Sox4* was found to correlate with stages of cancer and reduction in the survival rate, making it a potential biomarker for aggressive bladder cancer phenotype (93).

Recent research has incorporated circular RNA to develop functional biomarkers for bladder cancer. The involvement of circRNA in gene transcription as well as translation suggests the potential participation of circRNA in the process of disease progression including cancer (94). Results from Li et al(95) RNA-Seq study demonstrated downregulation of circular RNA (circHIPK3) in cell lines and tissues of bladder cancer. CircHIPK3 can sponge miR-558 which in turn suppresses heparanase (*HPSE*) expression as demonstrated mechanistically. The circHIPK3 overexpression was found to halt the bladder cell angiogenesis, migration, and invasion in vitro and inhibit the metastasis and growth of bladder cancer in vivo, suggesting a potential biomarker for bladder cancer therapy (95). Similarly, a study by Dong et al, (94) utilized a novel circRNA and circACVR2A of which their overexpression is linked with migration, proliferation and invasion of bladder cancer cells (94). It was demonstrated that circACVR2A is able to interact directly with miR-626, thus acting as a miRNA sponge which in turn regulates the expression of *EYA4* (94). Despite the above findings and the potential of several genes as bladder cancer biomarkers, to date, there have been no available prognostic and diagnostic biomarkers that have successfully been translated clinically.

### 4.3 Identified molecular pathways in bladder cancer

Modification of different molecular pathways and changes in living organisms such as mutations, alterations in gene control and epigenetic alterations are the driving forces for malignancy and its development (96). Alterations of signalling pathways, metabolic pathways, cytoskeleton and DNA repair pathways have been identified to be associated with tumorigenesis and progression of different types of cancers (97). Moreover, signalling pathways that control cell growth, proliferation, cell specialization and apoptosis, when damaged could lead to carcinogenesis and disease progression (97-99). Activation of focal adhesion and MAPK signalling pathways and subsequent dysregulating genetic processes are further uncovered in the development of bladder cancer (100).

Data generated by RNA sequencing and bioinformatics analyses revealed that 17 differentially expressed genes were downregulated in 5637 cell line whereas 44 were up-regulated in comparison to T24 (46). Similarly, down-regulation of WNT9A and WNT10A was confirmed in both cell lines which have been found to be associated with alteration of Wnt signalling pathways that contributes to bladder cancer initiation and development(1, 9) Using RNA-Seq, the most significantly enriched pathway involved in the development of bladder cancer was found to be the cell adhesion molecules (CAMs) pathway (FDR= 2.67E-08) (101). Further enrichment of bladder cancer pathway, focal adhesion, and extracellular matrix (ECM) receptor interaction pathways revealed that genes such as *PTPRF*, *VEGFA* and *CLDN7* are involved (or: are implicated) in all the pathways including bladder cancer (101). The upregulation of genes such as *ITGA*, *F3*, *ANXA1* has been reported in cancer-related pathways associated with cell growth, cellular cycle and apoptosis. However, downregulation of *GRB7*, *VEGP* genes was also noted in the same pathways (102).

Furthermore, upregulation of IFN-γ, angiogenesis, and inflammatory pathways was observed in NEURAL, mesenchymal-like (MES), and squamous-cell carcinoma-like (SCC) (103). Nevertheless, downregulation of several DEGs has been reported in immune activation pathways like NF-κβ signalling pathway, MAPK signalling pathway and PI3K-Akt signalling pathway (104). Interleukin-10 production, lymphocyte chemotaxis and aberrant IFN-γ, NF-κB and ERK signalling networks are the major pathways identified in the immune transcriptome related to programmed death-ligand 1 (PD- L1) inhibitors status and may participate in the evasion of immunity in high- grade muscle invasive urothelial carcinoma of the bladder (HGUC) (105). Similarly, PIK3/AKT/MTOR pathway is activated by ERBB2 and FGFR3 (types of receptor tyrosine kinases) and is found to control essential carcinogenesis stages and progression of tumour (106). In a different study, key signalling pathways including vascular endothelial growth factor (VEGF/VEGFR) pathway and PI3K-Akt-mTOR pathway were identified and have been linked to promote muscle-invasive bladder carcinoma. Also, through activation of Janus kinase-signal transducer and activator of transcription (JAK/STAT) signalling pathway, carcinogenesis of urothelial carcinoma was improved by Insulin-like growth factor binding protein 4–1 (IGFBP4–1). Upregulation of IGFBP4–1 in bladder tissue might play a significant role in the initiation and progression of bladder cancer (71). Similarly, L-type amino acid transporter 1 (LAT1), which transports leucine (essential amino acid), controls mammalian Target of Rapamycin (mTOR) signalling pathway and has been linked to the initiation and progression of urothelial carcinoma (107).

Interestingly, RNA-sequencing has shown significant upregulation (fold change > 2.0) of 1793 genes, on the other hand, downregulation of 1759 genes was recorded after knockdown of Methyltransferase-like 3 (METTL3), suggesting that methylation of METTL3 may halt the process of $N^6$-methyladenosine (m6A) methylation which could lead to the progression of urothelial carcinoma (108). The functional enrichment analysis revealed that strong (significant) negative correlation was observed in MYC oncogene and TNF-α/NF-κB target gene pathways, while strong positive correlation was noticed in other signalling pathways in relation to METTL3 knockdown(108).

In addition, increased activation of RAS pathway was observed in all BC159-T samples than TCGA-urothelial bladder carcinoma (TCGA- BLCA) samples (79). The alterations of DNA in urothelial carcinoma are controlled by the activation of the Ras–MEK–ERK and PI3

kinase–AKT–mTOR pathways either by tumour suppressor genes or oncogenes and thus, RB1-dependent G1-S cell cycle restriction point, anabolic metabolism, and cell survival RB1-dependent G1/S checkpoint, cell continuity (survival) and building aspect of metabolism (anabolism) regulate carcinogenesis through these pathways (109). In about 20% of muscle-invasive bladder neoplasm, mutation in fibroblast growth factor (FGFR3) was activated, triggering activation of receptor that encourages profuse growth through downstream activation of ERKs (109, 110). Genetically altered FGFR3 (mutant FGFR3) has been reported to activate RAS-MAPK pathway which led to abnormal proliferation of cancer cells (106).

Several attempts have been put in place to avert proliferation of cancer cells by targeting the specific molecular pathways for the development of bladder cancer. For example, anticancer drugs like infigratinib and dasatinib were used to target Erk1/2 and Src signalling pathways and thus overcoming drug  resistance observed when using other drugs (111). Similarly, three combination of anticancer drugs (romidepsin + gemcitabine + cisplatin) have been used to target the ERK pathway by increasing the level of reactive oxygen species (ROS) which trigger cysteine-aspertic proteases (caspases) that play a vital role in cell apoptosis (112). Therefore, understanding integrated metabolomic and transcriptomic pathways related to bladder cancer could offer alternative  ways for the treatment of bladder cancer (113).

### 4.4 Bladder cancer classification

Bladder cancer was earlier annotated with mutation recurrence in 32 genes (114) and characterized with signature genetic metabolism predispositions that include N-acetyl transferase editosome due to “apolipoprotein B mRNA editing enzyme, catalytic polypeptide-like” (APOBEC)-cytidine deaminases (10). Profiling of 58-genes and genes fusion of 784 genes have identified APOBEC signature mutations (115), that are highest in bladder cancer due to overexpression of APOBEC3B genes than in all solid tumours(109). Accurate association of bladder cancer histological variants pathological features and specific molecular mutational pattern might provide promising target for developing therapy (116). So far, the European Association of Urology has provided guidelines on identifying evidence-based prognosis and progression(115) as different morphological subtype has different clinical course and response to treatment (117). Lin and colleagues(102) recommend the use of radiomics and transcriptomics in further identification of prognostic signature markers for the different variants of bladder cancer.

The two main types of bladder cancer include the muscle-invasive bladder cancer (MIBC) and non-muscle-invasive bladder cancer (NMIBCs). The remarkable difference between them is that the NMIBCs are of low-grade tumour(103, 115), accounting for about 75% of all diagnosed cases(10, 103) and has been associated with a relative stable genomic *FGFR* mutation(10, 103) and homozygous deletion of *CDKN2A* (118). Gene expression profiling has identified 3 subtypes of NMIBCs from a luminal to basal type marker progression shift from class 1 to class 3 (119). Other type of bladder cancer is the muscle-invasive bladder cancer (MIBC), although less frequently diagnosed, it is genetically predisposed as unstable and divergent type(109, 119). MIBC progress from a flat lesion and gradually acquires gene mutations that promote cell proliferation and survival(120) and are associated with aneuploidy *TP53* mutations (10, 103). Other mutations observed with MIBCs include p53 expression stability, upregulated cytokeratin 20 (*CK20*) and *HER2*/neu genes and downregulated *PTEN* expression that is associated with upregulation of phosphatidylinositol 3-kinase (PI3K) pathway (10). Additionally, the common RB1 deletion and 6P amplification are associated with MIBCs development(118) with an unfortunate less than 50% 5-year survival  (103, 117). However, the NMIBCs exhibits a papillary growth pattern of low malignancy potential, and require an expensive clinical management which includes local resection due to high tendency of recurrence (115, 120).

Nevertheless, the earlier anatomical based classification of bladder tumour (NMIBC; Tis, Ta &T1 and MIB; T2, T3 & T4) had limitation in the understanding of the diverse prognosis within same tumour type (121). Several studies have speculated that tumour heterogeneity might be associated with diagnosis, metastasis and pre-existing treatment resistance especially in the case of MIBCs (121). Therefore, for an optimized patient precision therapeutic management, identification of an accurate histopathology (116, 122) and specific transcriptomic biomarkers are encouraged to be adopted (123).[3] Based on this, bladder cancer has been classified into urothelial carcinoma (UC), urothelial carcinoma with variant (UCV), squamous cell carcinoma (SCC), and adenocarcinoma (AD) subtypes (124, 125). Transcriptomics and other molecular techniques have led to the identification of urothelial carcinoma (UC) that account for 75% (116)  or even up to 90%(126) of the bladder cancer. Whereas the remaining 25%, comprises of other histological variants known as the nonurothelial cancer (116). Invasive UC have a tendency to progress into divergent variants of nonurothelial tumours through progressive histological differentiation (126). Nonurothelial tumours are also known to consist of different range of rare variants based on the WHO 2016 classification that include the pure squamous, glandular, neuroendocrine, sarcomatoid, micropapillary, plasmacytoid, microcystic, clear cell and pure histological variants (116, 117). The pasmacytoid variant has been reported to have the worst overall survival mean (116, 124). In addition, other sub-populated tumours with either the pure squamous cell or glandular differentiation are shown not to be sensitive to adjuvant chemotherapy, unlike their counterpart that are populated with small cell carcinoma which are aggressive and sensitive to neo-adjuvant chemotherapy (124).

Furthermore, UC is more prevalent in developed countries like Japan, North America, and some Western countries (122), while squamous cell carcinoma (SCC) and adenocarcinoma have a higher prevalence in countries like Egypt (125). Zhang et al(101) used high-throughput transcriptomic profiling in cisplatin resistant UCs, to unravel differential splicing in more than 300 genes with the specific dysregulation in five candidate genes were validated with real time PCR. These including upregulation of *CDH1*, *VEGFA*, *PTPRF*, *CLDN7* genes and downregulation of *MMP2* gene that are all together involved in cell adhesion molecules (CAMs), focal adhesion and bladder cancer progression pathways (101). In rare UC cases, fusion transcripts associated with chromosomal region rearrangements (17q25, 15q26.3 and 1p36.22) have been identified as *SET9/CYHR*, *IGF1R/TTC23*, *SYT8/TNN12 and CASZ1/DFFA* transcripts respectively (47). Similarly, about 25-30% of UC of the bladder possess features of histological variant which are predominantly populated with SCC (125),[125] which are enriched in *MAPK* signalling pathway (121). Although SCC of the bladder represents a small fraction (2%-5% in the United State) of prevalence in the western countries based on the National Cancer Database (NCDB), it attributes the worse prognosis both in terms of stage-for-stage, grade, relapse and overall survival to the UC (116, 127). Both the UC and SCC are derived from UCV and have been demonstrated to have identical driving genes but completely different transcriptional profiles (128). On the other hand, SCC has also been attributed to be less responsive to chemotherapy, thus its standard pre-existing treatment is still radical cystectomy (RC) which might provide the maximal 5-year survival (129). Unlike UC, SCC is shown to have a negative expression of immune inhibitor biomarkers such as PD-L1, thus its higher grade and aggressive association. Owyong *et al(129)* suggests the increase in the expression of PD-L1 might improve the immune response associated with treatment outcome. Primary adenocarcinoma represents 0.5-2% of the all diagnosed bladder cancer, like SCC it has a poor prognosis and has a weak TM immunoexpression score than UC (130).

On the other hand, small cell carcinoma (SmCC) is a rare bladder cancer variant (less than 1% of all bladder diagnosis) and belongs to the neuroendocrine carcinoma (NEBC) family, usually presented at a late stage (126). Neuroendocrine carcinoma is somewhat similar and sometimes overlaps with UC. Although relatively rare (0.5-1.2% of new cases), it is highly lethal with a 1-year survival progression (126, 131). Small cell carcinoma of the bladder is associated with a depletion of *CCND1* and an amplification of *CDKN2A* genes whereas in UC, *CDKN2A* gene is depleted while *CCND1* gene is amplified (131). This variant has neuronal marker signatures including upregulation of *NESTIN*, *TUBB2B, PEG10* gene with low or no expression of basal or luminal biomarkers (131, 132). Another identified novel marker of SmCC is dysregulated PVT1-ERBB2 affecting ERBB2 gene expression which compliments the general MIBC signature markers; TP53 and RB1 depletion (126, 133), and are evident in their histological features (131). Although SmCC/ NEBC may express basal or luminal biomarkers, a few of them have a basal molecular feature which might explain their positive response to chemotherapy (126, 132) or radiotherapy in improving survival (132). Study by Shen and others, shows that NEBC is sensitive to P13k inhibitor (GDC-0941) and FGFR inhibitor (NVP-BGJ398). While Wang et al(126) recommended utilizing the common RB1 gene depletion as a potential therapeutic target.

**5.0 Challenges and limitations of RNA-Seq method**

RNA sequencing (RNA-Seq) has become an essential tool for analysing differential gene expression (DGE) utilized in characterizing specific tissues (134). This development has given room for deeper insight into the complexity of the protein-coding transcriptome than previously understood. Some studies have preferred RNA-Seq over the earlier developed high-throughput RNA analysis by microarrays due to its numerous advantages (17). One of such promising opportunities is the detection of gene fusions and differential expression of transcripts known to cause disease. Lee et al(52) reported that dysregulation of long non-coding RNAs had been associated with development of some diseases such as cancer, myocardial infarction and diabetes. Other advantages include increased dynamic range of expression, measurement of focal changes (such as single nucleotide variants (SNVs), insertions and deletions), detection of rare and novel transcript isoforms, splice variants and chimeric gene fusions (that include genes and transcripts hitherto not identified) (20).

Despite the merits of the RNA-Seq, some limitations especially associated with the library preparation protocols can cause biases and overvaluation of results(27, 135). Because sequencing is sensitive to the quantity of transcripts, this can make abundant mRNAs to be overly represented in RNA-Seq libraries and can have influence on the majority of the reads. These mRNAs are evaluated with low stochastic variability between samples and can be found significant by differential expression analysis (DEA). Conversely, transcripts low in abundance receive few reads, which can subject them to noise and reduce their chances of being selected (27).

Another factor that influences the transcript detection in RNA-Seq experiments is the length of the transcript itself, because a longer transcript has a higher possibility of being detected in the library, it will be considered significant after DEA which can subsequently affect the functional annotation of significant genes (136, 137). Excitingly, evidences have proven that microarrays can outperform RNA-Seq in detecting small non-coding RNAs like microRNAs. Another limitation of the RNA-Seq is the errors that arise from the quality of the RNA, as significant percentage of the total RNA come from ribosomal RNA (rRNA) while a very low percentage come from the mRNA, as a result, special considerations must be made in the methods in order to either enrich the mRNA (polyA selection) or reduce the rRNA levels (17). It is possible to avoid some of these issues with proper study design. However, bias and inconsistency associated with adapter ligation, cDNA synthesis, and amplification could be primarily dependent on library preparation and pre-processing procedures (138). Therefore, standardized procedures for addressing the limitations of the RNA-Seq and ascertaining the accuracy, reproducibility and precision in a clinically important settings are pertinent to enable the adoption of RNA-Seq tests.

## 6.0 Conclusion and future perspective

Urothelial cell carcinoma (UCC) is among the prime causes of malignancy-associated deaths globally and its aetiology is poorly understood. The bladder cancer progression is a multifaceted process that is extremely heterogeneous and complicated. Thus, requiring the use of advanced transcriptome technology like RNA-sequencing (RNA-Seq) to better understand the disease phenotypes. The RNA-Seq studies of bladder cancer reviewed in this paper reveal genome-wide changes among different classes of bladder cancer and its pathogenicity that includes; dysregulated genes and pathways. Moreover, the possibilities to better understand the molecular mechanism underlying UCC progression has been successful through application of RNA-sequencing as this helps in understanding pathogenesis, discovering biomarkers and uncovering the mechanism of drug resistance in bladder cancer.

Despite the progressive improvement provided by RNA-Seq technologies in understanding the molecular basis for UCC pathogenesis, there is still a lot to be done to prevent recurrence and further metastasis of the malignant cells, and to reduce cancer-related death due to UCC. A good number of dysregulated genes and biomarkers have been implicated in bladder cancer, and the uses of these genes clinically have provided controversial results. Hence, there has not been a single prognostic and diagnostic biomarker that are successfully translated into clinical application. This opens up an avenue for a potential research space to fully understand bladder cancer. To address the current limitations, there is a need to have a standard pipeline and integrated multiple experiments analyses as a reference point for all studies due to the high recurrence nature of UCC. The existence of molecular subsets of cellular signatures and microenvironment promotes measures, which could avoid therapeutic failures exposed by RNA-Seq study. Moreover, acceptability of RNA-Seq regarding gene expression in UCC led to existence of incompatibility due to RNA-Seq techniques and variability in platform application.

In summary, RNA-Seq is regarded as the golden standard for large scale high-throughput genome-wide and gene expression studies. It has provided us an unknown insight into the transcriptional changes in bladder cancer diagnosis and management. As the cost of sequencing is gradually decreasing and more precision is being obtained in most of the computational tools, RNA-Seq technology will continue to be applied in studying bladder cancer transcriptomics and disease state. Consequently, an in-depth exploration of the complex and stochastic genetic nature of bladder cancer could possibly result in a novel biomarker discovery and identifying new therapeutic strategies. Moreover, single cell sequencing (scRNA-Seq) may offer a better option for comprehensive understanding of UCC development and recurrence and could effectively help in improving bladder cancer management.

## Authors contributions

**Umar Ahmad**: Conceptualisation, Methodology, Software, Investigation, Resources, Writing- Original Draft, Writing- Review & Editing; Visualization; Supervision. **Buhari Ibrahim**: Data curation, Writing - original draft, Writing- Review & Editing. **Mustapha Mohammed**: Data curation, Writing - original draft. **Ahmed Faris Aldoghachi**: Writing - Original Draft. **Mahmood Usman**: Writing - Original Draft. **Abdulbasit Haliru Yakubu**: Data Curation; Writing - Original Draft. **Abubakar Sadiq Tanko**: Writing - original draft. **Bobbo Khadijat Abubakar**: Writing - original draft. **Usman Adamu Garkuwa**: Writing - original draft. **Abdullahi Adamu Faggo**: Writing - original draft. **Sagir Mustapha**: Writing - original draft. **Mahmoud Al-masaeed**: Writing- Review & Editing, Funding acquisition; **Syahril Abdullah**: Supervision, Resources, Funding acquisition. **Yong Yoke Keong**: Methodology, Writing- Review & Editing; **Abhi Veerakumarasivam**: Supervision, Resources, Funding acquisition.

## Conflict of interest

Declarations of interest: none

## Acknowledgements

Support for title page creation and format was provided by AuthorArranger, a tool developed at the National Cancer Institute (NCI), United States. All figures in this manuscript are designed and created with BioRender.com..

## Abbreviations

| | |
|---|---|
| AD | Adenocarcinoma |
| CAMs | Cell Adhesion Molecules |
| cDNA | Complimentary Deoxynucleic Acid |
| circRNA | Circular RNA |
| DEA | Differential Expression Analysis |
| DEGs | Differentially expressed genes |
| DNA | Deoxynucleic Acid |
| ECM | Extracellular Matrix |
| GUI | Graphical User Interface |
| HGUC | High-grade Urothelial Carcinoma |
| lncRNA | Long non-coding RNA |
| MES | Mesenchymal-like |
| MIBC | Muscle Invasive Bladder Cancer |
| miRNA | microRNA |
| mTOR | Mammalian Target of Rapamycin |
| NCDB | National Cancer Database |
| NCI | National Cancer Institute |
| NEBC | Neuroendocrine Carcinoma |
| NGS | Next-generation Sequencing |
| NK | Natural Killer |
| NMIBC | Non-muscle Invasive Bladder Cancer |
| Nt | Nucleotide |
| PacBio | Pacific Biosciences |
| PCR | Polymerase Chain Reaction |
| QC | Quality Control |
| RC | Radical Cystectomy |
| RIN | RNA Integrity Number |
| RNA | Ribonucleic Acid |
| RNA-Seq | RNA Sequencing |
| ROS | Reactive Oxygen Species |
| RPKM | Reads Per Kilobase of Exon Model Per Million Reads |
| rRNA | Ribosomal RNA |
| RT | Reverse Transcriptase |
| SCC | Squamous Cell Carcinoma |
| scRNA-Seq | Single Cell RNA Sequencing |
| SMRT | Single-Molecule Real-Time |
| SmSCC | Small Cell Carcinoma |
| SNVs | Single Nucleotide Variants |
| SOLiD | Sequencing by Oligonucleotide Ligation and Detection |
| STAR | Spliced Transcript Alignment to a Reference |
| UA | Umar Ahmad |
| UC | Urothelial Carcinoma |
| UCC | Urothelial Cell Carcinoma |
| UCSC | University of California Senta Crus |
| UCV | Urothelial Carcinoma with Variant |

## Figure Legends

**Figure 1: Overview of RNA-sequencing on bladder cancer**. RNA is first extracted from the bladder cancer tissues or cells samples followed by the cDNA synthesis through reverse transcription. Then, the cDNAs are amplified for library preparation before subjecting them to next-generation sequencing (NGS) to generate an output of raw FASTQ data files (reads) that could be used for downstream computational and bioinformatics analyses.

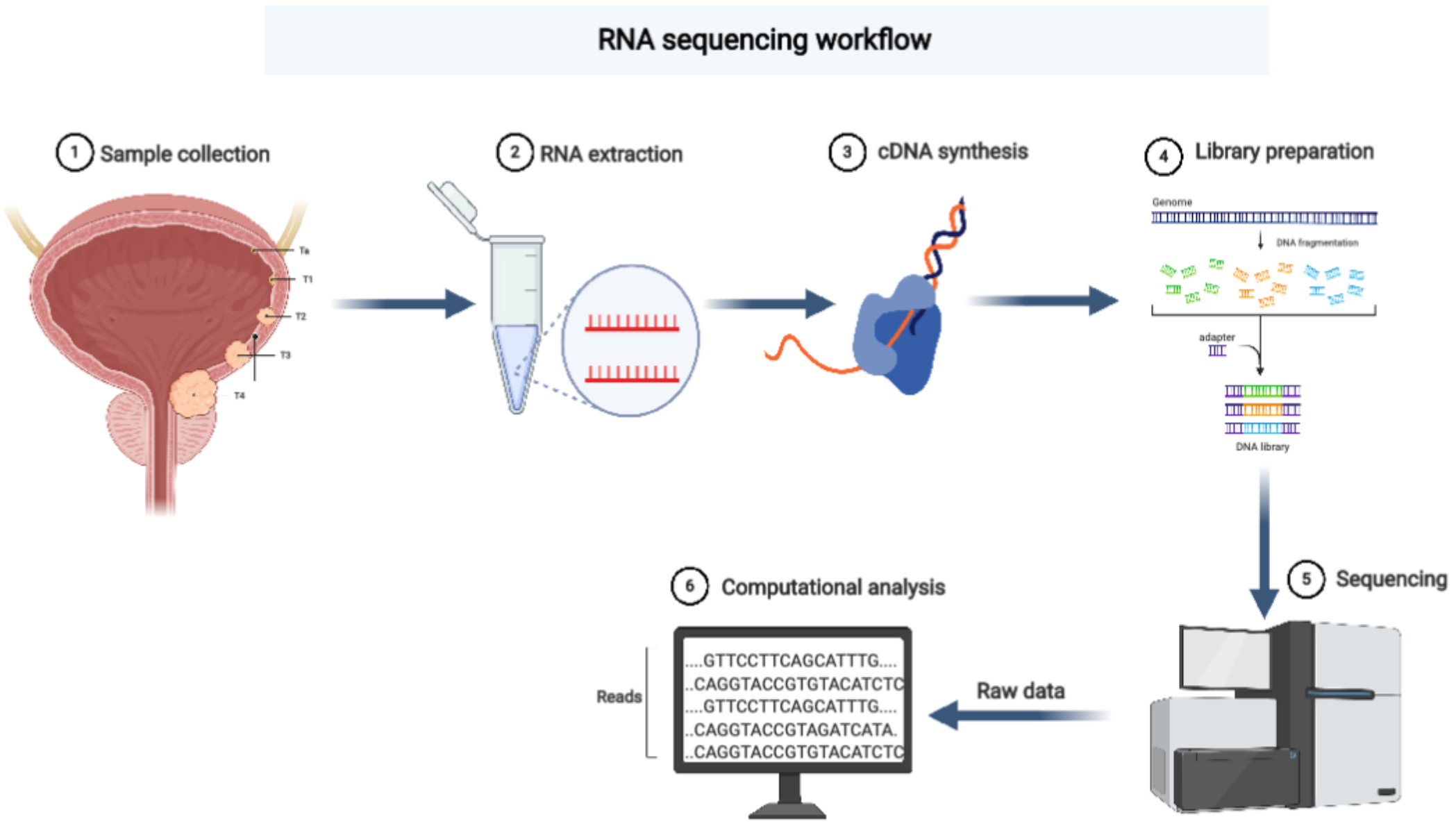

**Figure 2: General RNA-Seq analysis pipeline**. A highlight of the stepwise procedure involved in RNA-Seq data analysis. A typical RNA-Seq analysis workflow consists these seven steps with each step having some bioinformatics analysis ~~to perform~~ before going downstream to the target objectives of the study.

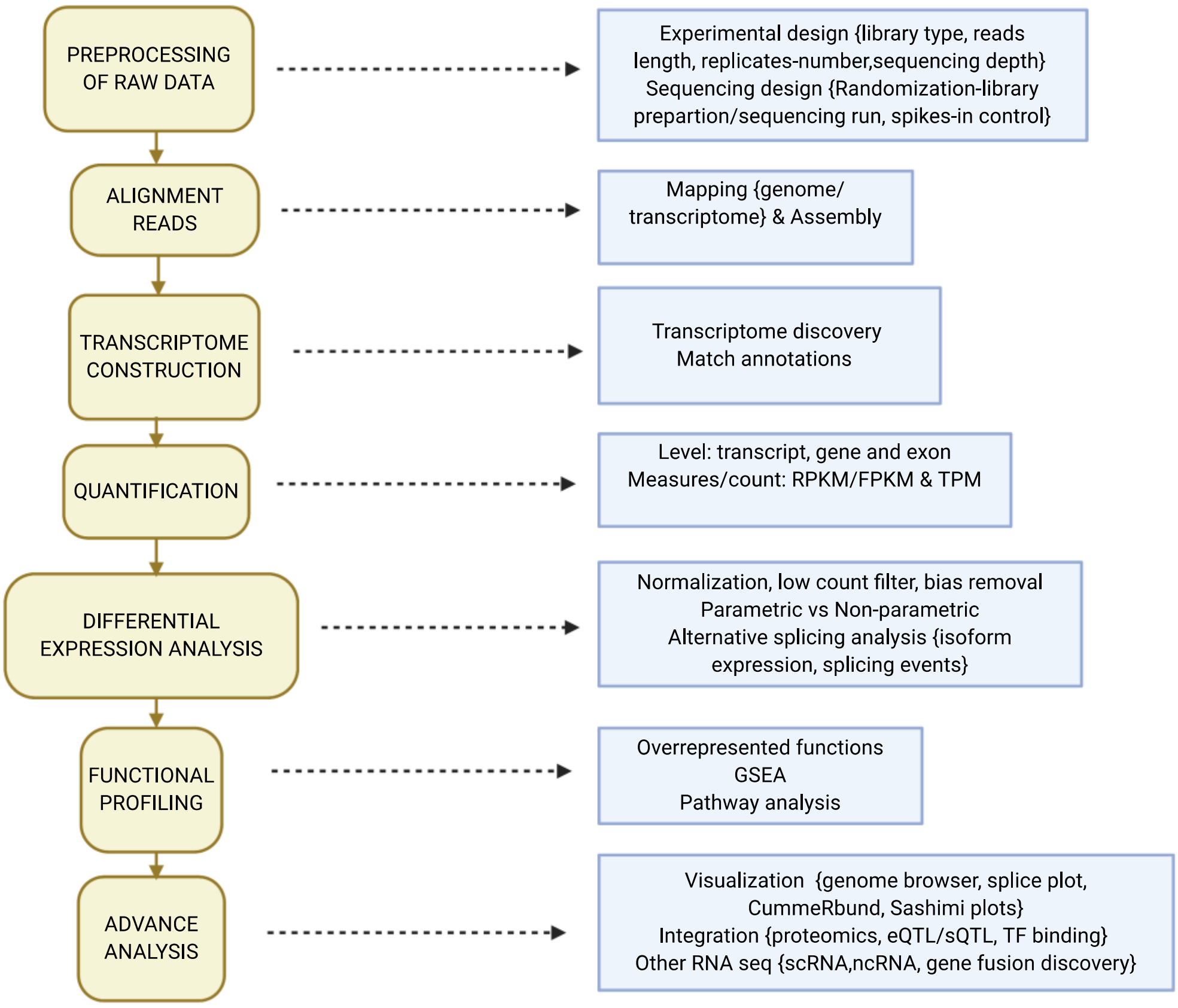

## Table 1: List of platforms for RNA-Seq analysis

| S/N | Software | Platform | Link (URL) | Description | Reference |
|---|---|---|---|---|---|
| 1 | Python, R | UTAP | https://utap.readthedocs.io | UTAP (User-friendly Transcriptome Analysis Pipeline) is an open source, web-based intuitive and scalable transcriptome pipeline that executes the full process, starting from sequences (RNA-Seq and bulk MARS-Seq), ending with sets of differentially expressed genes and sophisticated reports, and requiring minimal user expertise. | Kohen et al., 2019 |
| 2 | Python | RASflow | https://github.com/zhxiaokang/RASflow. | RASflow (RNA-Seq Analysis Snakemake Workflow) is a lightweight and easy-to-manage RNA-Seq analysis workflow. It includes the complete workflow for RNA-Seq analysis, starting with QC of the raw FASTQ files, going through optional trimming, alignment and feature counting (if the reads are mapped to a genome), pseudo alignment (if transcriptome is used as mapping reference), gene- or transcript- level DEA, and visualization of the output from DEA. | Zhang and Jonassen, 2020. |
| 3 | R | ARMOR | | ARMOR (Automated Reproducible MOdular RNA-Seq) is a Snakemake workflow, aimed at performing a typical RNA-seq workflow in a reproducible, automated, and partially contained manner. It is implemented such that alternative or similar analysis can be added or removed. | Orjuela et al., 2019 |
| 4 | R, Python, Unix, C++, java, Perl | VIPER | https://bitbucket.org/cfce/viper/ | VIPER (Visualization Pipeline for RNA-Seq) is a comprehensive RNA-Seq analysis tool built using snakemake which allows for ease of use, optimal speed, and a highly modular code that can be further added onto and customized by experienced users. VIPER combines the use of several dozen RNA-Seq tools, suites, and packages to create a complete pipeline that takes RNA-Seq analysis from raw sequencing data all the way through alignment, quality control, unsupervised analysis, differential expression, and downstream pathway analysis. | Cornwell et al., 2018 |
| 5 | web | Biojupies | http://biojupies.cloud. | BioJupies is a web application that enables the automated creation, storage, and deployment of Jupyter Notebooks containing RNA-Seq data analyses. Through an intuitive | Torre et al., 2018. |
| | | | | interface, novice users can rapidly generate tailored reports to analyse and visualize their own raw sequencing files, gene expression tables, or fetch data from >9,000 published studies containing >300,000 pre-processed RNA-Seq samples. | |
| 6 | Perl, R | hppRNA | https://sourceforge.net/projects/hpprna/. | hppRNA package is dedicated to the RNA-Seq analysis for a large number of samples simultaneously from the very beginning to the very end, which is formulated in Snakemake pipeline management system. It starts from fastq files and will produce gene/isoform expression matrix, differentially-expressed-genes, sample clusters as well as detection of SNP and fusion genes by combination of the state-of-the-art software. | Wang, 2018 |
| 7 | Python | RNA Cocktail | http://bioinform.github.io/rnacocktail/. | The RNACocktail pipeline is composed of high-accuracy tools for different steps of RNA-Seq analysis. It performs a broad-spectrum RNA-Seq analysis on both short- and long-read technologies to enable meaningful insights from transcriptomic data. It was developed after analysing a variety of RNA-Seq samples (ranging from germline, cancer to stem cell datasets) and technologies using a multitude of tool combinations to determine a pipeline which is comprehensive, fast and accurate | Sahraeian et al., 2017 |
| 8 | Python | aRNApipe | https://github.com/HudsonAlpha/aRNAPipe | aRNApipe is a project-oriented pipeline for processing of RNA-seq data in high performance cluster environments. The provided framework is highly modular and has been designed to be deployed on HPC environments using IBM Platform LSF, although it can be easily migrated to any other workload manager. | Alonso et al., 2017 |
| 9 | Python, R | UTAP | https://utap.readthedocs.io | BISR-RNAseq (Bioinformatics Shared Resource Group-RNAseq) is a consistent workflow that allows for the analysis (alignment, QC, gene-wise counts generation) of raw RNAseq data and seamless integration of quality analysis and differential expression results into a configurable R shiny web application. | Kohen et al., 2019 |
| 10 | Python, R, Shell | RASflow | https://github.com/zhxiaokang/RASflow | TRAPLINE (Transparent, Reproducible and Automated PipeLINE) supports NGS-based research by providing a workflow that requires no bioinformatics skills, decreases the processing time of the analysis, and works in the cloud | Zhang and Jonassen, 2020 |
| 11 | Python, R & Bash | BISR-RNAseq | https://github.com/MPiet11/BISR-RNAseq | The docker4seq package was developed to facilitate the use of computing demanding applications in the field of NGS data analysis. It uses docker containers that embed demanding computing tasks (e.g., short reads mapping) into isolated containers. | Gadepalli et al., 2019 |
| 12 | web | TRAPLINE | https://usegalaxy.org/u/mwolfien/p/trapline---manual | QuickRNASeq is an open-source based pipeline for large scale RNA-Seq data analysis. It takes advantage of parallel computing resources, a careful selection of previously published algorithms for RNA-Seq read mapping, counting and quality control, and a three-stage strategy to build a fully automated workflow. | Wolfien et al., 2016 |
| 13 | R | Docker4seq | http://reproducible-bioinformatics.org | IRIS-EDA (Interactive RNA-Seq Interpretation System for Expression Data Analysis) provides a user-friendly interactive platform to analyse gene expression data comprehensively and to generate interactive summary visualizations readily. In contrast to other analysis platforms, IRIS-EDA provides the user with a more comprehensive and multi-level analysis platform. | Kulkarn et al., 2018. |
| 14 | Bash scripting, Perl, R, JavaScript | QuickRNAseq | http://quickrnaseq.sourceforge.net | bcbioRNASeq is a Bioconductor package that provides ready-to-render templates, objects and wrapper functions to post-process bcbio RNA sequencing output data. It helps automate the generation of high-level RNA-Seq reports, facilitating the quality control analyses, identification of differentially expressed genes and functional enrichment analyses. | He et al., 2018 |
| 15 | web | IRIS-EDA | http://bmbl.sdstate.edu/IRIS/. | START (Shiny Transcriptome Analysis Resource Tool) provides researchers with increased flexibility to easily upload and visualize RNA-Seq data. The App visualizes data in multiple ways that will be useful for scientists to understand their data. Critical to facilitating data sharing capabilities, the App can be utilized within a web browser environment for easy access as well as enabling seamless sharing of data between collaborators. | Monier et al., 2019 |
| 16 | R | bcbioRNASeq | https://github.com/hbc/bcbioRNASeq | iDEP (integrated Differential Expression and Pathway analysis) enables users to conduct in-depth bioinformatics analysis of transcriptomic data through a GUI. The two use cases demonstrated that it can help pinpoint molecular pathways from large genomic datasets, thus eliminating | Steinbaugh et al., 2018 |
| | | | | some barriers for modern biologists. | |
| 17 | R | START | https://kcvi.shinyapps.io/START | DEApp (Differential Expression App) interactive and dynamic web application for differential expression analysis of count based NGS data. It enables models selection, parameter tuning, cross validation and visualization of results in a user-friendly interface. | Nelson et al., 2017 |
| 18 | R | iDEP | http://ge-lab.org/idep/ | GENAVi (Gene Expression Normalization Analysis and Visualization) provides a user-friendly interface for normalization and differential expression analysis (DEA) of human or mouse feature count level RNA-Seq data. It is a GUI based tool that combines Bioconductor packages in a format for scientists without bioinformatics expertise. | Ge et al., 2018 |
| 19 | R | DEApp | https://yanli.shinyapps.io/DEApp | TCC-GUI (Graphical User Interface for TCC) is a browser-based application for DE analysis of RNA-Seq data. It enables non-R users to perform the TCC package without installation. In addition to the functionalities originally implemented in TCC, TCC-GUI provides plenty of interactive visualization functions. The powerful in-built functions would also be satisfactory for experienced R users. | Li and Andrade, 2017 |
| 20 | R | GENAVi | https://junkdnalab.shinyapps.io/GENAVi/ | BEAVR (Browser-based tool for the Exploration and Visualization of RNA-Seq data) is an easy-to-use tool that facilitates interactive analysis and exploration of RNA-Seq data. It is developed in R and uses DESeq2 as its engine for differential gene expression (DGE) analysis, but assumes users have no prior knowledge of R or DESeq2. BEAVR allows researchers to easily obtain a table of differentially expressed genes with statistical testing and then visualize the results in a series of graphs, plots and heatmaps. | Rayes et al., 2019 |
| 21 | Web, R | TCC-GUI | https://infinityloop.shinyapps.io/TCC-GUI/ | iSeq, an R-based Web server, for RNA-Seq data analysis and visualization. iSeq is a streamlined Web-based R application under the Shiny framework, featuring a simple user interface and multiple data analysis modules. Users without programming and statistical skills can analyse their RNA-Seq data and construct publication-level graphs through a standardized yet customizable analytical pipeline. | Su et al., 2019 |
| 22 | R | BEAVR | https://github.com/developerpiru/BEAVR and https://hub.dock | DaMiRseq package is a structured and convenient workflow to effectively identify transcriptional biomarkers and exploit | Perampalam and Dick,2020 |
| | | | er.com/r/pirunthan/beavr | them for classification purposes. | |
| 23 | R | iSeq | http://iseq.cbi.pku.edu.cn. | DEsubs is a network-based systems biology R package that extracts disease-perturbed sub pathways within a pathway network as recorded by RNA-Seq experiments. It contains an extensive and customized framework with a broad range of operation modes at all stages of the sub pathway analysis, enabling a case-specific approach. | Zhang et al., 2018 |
| 24 | R | DaMiRseq | https://bioconductor.org/packages/DaMiRseq/ | CANEapp (Comprehensive automated Analysis of Next-generation sequencing Experiments App) is a unique suite that combines a Graphical User Interface (GUI) and an automated server-side analysis pipeline that is platform-independent, making it suitable for any server architecture. | Chiesa et al., 2018 |
| 25 | R | DEsubs | http://bioconductor.org/packages/DEsubs/. | DiCoExpress is an R script-based tool allowing users to perform a full RNA-Seq analysis from quality controls to co-expression analysis through differential analysis based on contrasts inside generalized linear models. DiCoExpress focuses on the statistical modelling of gene expression according to the experimental design and facilitates the data analysis leading the biological interpretation of the results. | Vrahatis et al., 2016 |
| 26 | Python, Java | CANEapp | http://psychiatry.med.miami.edu/research/laboratory-of-translational-rna-genomics/CANE-app | IRIS-DGE (Integrated RNA-Seq Data Analysis and Interpretation System for Differential Gene Expression) is a server-based DGE analysis tool developed using Shiny. It provides a straightforward, user-friendly platform for performing comprehensive DGE analysis, and crucial analyses that help design hypotheses and to determine key genomic features | Velmeshev et al., 2016 |
| 27 | R | DicoExpress | https://forgemia.inra.fr/GNet/dicoexpress | SPARTA (Simple Program for Automated reference-based bacterial RNA-Seq Transcriptome Analysis) is a reference-based bacterial RNA-Seq analysis workflow application for single-end Illumina reads. SPARTA is turnkey software that simplifies the process of analysing RNA-Seq data sets, making bacterial RNA-Seq analysis a routine process that can be undertaken on a personal computer or in the classroom. | Lambert et al., 2020. |
| 28 | R | IRIS-DGE | http://bmbl.sdstate.edu/IRIS/ | RAP (RNA-Seq Analysis Pipeline) is a web application implementing a fully automated analysis workflow, designed to integrate in-house developed scripts as well as open-source analysis tools into one pipeline. | Monier et al., 2018. |
| 29 | Python | SPARTA | http://sparta.readthedocs.org | Shiny-Seq provides a guided and easy to use comprehensive RNA-Seq data analysis pipeline. It has many features such as batch effect estimation and removal, quality check with several visualization options, enrichment analysis with multiple biological databases, identification of patterns using advanced methods such as weighted gene co-expression network analysis, summarizing analysis as PowerPoint presentation and all results as tables via a one-click feature | Johnson et al., 2016 |
| 30 | web | RAP | http://bioinformatics.cineca.it/rap/. | The Cancer Genome Atlas (TCGA) is a large-scale study that has catalogued genomic data accumulated for many different types of cancers, and includes mutations, copy number variation, mRNA and miRNA gene expression, and DNA methylation. Being publicly distributed, it has become a major resource for cancer researchers in target discovery and in the biological interpretation and assessment of the clinical impact of genes of interest. | D'Antonio et al., 2015 |
| 31 | R | Shiny-seq | https://szenitha.shinyapps.io/shiny-seq3/ | A multitude of large-scale studies, e.g., TCGA and GTEx, have recently generated an unprecedented volume of RNA-Seq data. The RNA-Seq expression data from different studies typically are not directly comparable, due to differences in sample and data processing and other batch effects. Here, we developed a pipeline that processes and unifies RNA-Seq data from different studies. Using the pipeline, we have processed data from the GTEx and TCGA and have successfully corrected for study-specific biases, allowing comparative analysis across studies. | Sundararajan et al., 2019 |
| 32 | R | TCGA RNA seq | https://github.com/srp33/TCGA_RNASeq_Clinical. | The Toil RNA-Seq workflow converts RNA sequencing data into gene- and transcript-level expression quantification. | Mumtahena et al., 2015 |
| 33 | R | RNAseqDB | https://github.com/mskcc/RNAseqDB/ | BioJupies is a web application that enables the automated creation, storage, and deployment of Jupyter Notebooks containing RNA-Seq data analyses. Through an intuitive interface, novice users can rapidly generate tailored reports to analyse and visualize their own raw sequencing files, gene expression tables, or fetch data from >9,000 published studies containing >300,000 pre-processed RNA-Seq samples. | Wang et al., 2018 |
| 34 | python | ToilRNA seq | https://github.com/BD2KGenomics/toil-rnaseq | hppRNA package is dedicated to the RNA-Seq analysis for a large number of samples simultaneously from the very | Vivian et al., 2017 |
| | | | | beginning to the very end, which is formulated in Snakemake pipeline management system. It starts from fastq files and will produce gene/isoform expression matrix, differentially-expressed-genes, sample clusters as well as detection of SNP and fusion genes by combination of the state-of-the-art software. | |

**Table 2: Summary of the current RNA-Seq studies on bladder cancer**

| S/N | Sample | Source | Grade | Library Preparation | Library Selection | Library layout | Sequencing Platform | Instrument Model | Average Mean reads Pair-end length (bp) | Reference |
|---|---|---|---|---|---|---|---|---|---|---|
| 1. | Tissue | UCC | Ta/T1/T2-4/CIS | ScriptSeq | cDNA | Paired-end | Illumina | Illumina HiSeq 2000 | 101 + 7 + 101 | Hedegaard et al. (2016) |
| 2. | Tissue | UCC | N/A | Ribo-Zero Magnetic Gold Kit (Illumina, USA) and the NEBNext® Ultra™ RNA Library Prep Ki | cDNA | Paired-end | Illumina | HiSeqX | N/A | Li et al. (2018) |
| 3. | Tissue | UCC | N/A | TruSeq Stranded mRNA Lib Prep Kit | cDNA | Paired-end | Illumina | NextSeq 500 | 72 | Maeda et al. (2018) |
| 4. | Cell | Urine | N/A | Ovation RNA Seq System V2 kit | cDNA | Paired-end/ Single | Illumina | Illumina HiSeq 2000 | N/A | Sin et al. (2017) |
| 5. | Cell | N/A | N/A | TruSeq Stranded Total RNA kit | cDNA | N/A | Illumina | Illumina HiSeq 2500 | N/A | Shen et al. (2015) |
| 6. | Tissue | N/A | N/A | RiboZero rRNA Removal Kit | cDNA | N/A | Illumina | HiSeq2000 | N/A | Li et al. (2017) |
| 7. | Tissue | UCC | N/A | TruSeq | cDNA | Pair-end | Illumina | Genome Analyzer IIx (GAIIx) | 380 | Zhang et al (2014) |
| 8. | Tissue/cell | UC | T1, T2–T4a, N0–3, M0–1 | DNA/RNA AllPrep kit (QIAGEN)/ mirVana miRNA isolation kit | cDNA, miRNA | N/A | Illumina | Illumina HiSeq 2500 | 76 | Robertson etal (2017) |